\documentclass[journal,comsoc]{IEEEtran}

\usepackage[T1]{fontenc}
\usepackage{xcolor}
\usepackage{graphicx}
\usepackage{xspace}
 
\usepackage{array}
\usepackage{makecell}

\usepackage{amsmath,amssymb}
\usepackage[cmintegrals]{newtxmath}
\usepackage{algorithm}
\usepackage{algpseudocode}

\usepackage[utf8]{inputenc}
\usepackage{amsmath}
\newtheorem{example}{Example}
\usepackage{pgfplots}
\pgfplotsset{compat=newest}

\pgfplotsset{
  every axis legend/.append style={
    legend cell align=left,
    align=left,
    font=\footnotesize
  }
}

\pgfplotsset{
  every axis plot/.append style={
      line width=1.5pt,
      mark size=2.5pt,
      mark options={solid,line width=0.75pt,fill=white!80!.}
    }
}

\pgfplotsset{
  every axis/.append style={
        scaled ticks = false, 
        tick label style={/pgf/number format/fixed},
    label style={font=\footnotesize},
        tick label style={font=\footnotesize}  
    }
}

\definecolor{UniformColor}{rgb}{0.00000,0.44700,0.74100}%
\definecolor{CapacityColor}{rgb}{0,0,0}%
\definecolor{CCDMInfColor}{rgb}{0,0.6,0}%
\definecolor{PBDMShortColor}{rgb}{0.85,0.325,0.098}%
\definecolor{PBDMMediumColor}{rgb}{0.929,0.694,0.125}%
\definecolor{PBDMLongColor}{rgb}{1,0,0}%

\pgfplotsset{Uniform/.style={dashed,color=UniformColor}}
\pgfplotsset{Capacity/.style={solid,line width=2pt,color=CapacityColor}}
\pgfplotsset{CCDMInf/.style={dotted,color=CCDMInfColor}}
\pgfplotsset{PBDMShort/.style={solid,color=PBDMShortColor, mark=diamond*}}
\pgfplotsset{PBDMMedium/.style={solid,color=PBDMMediumColor,mark=pentagon*}}
\pgfplotsset{PBDMLong/.style={solid,color=PBDMLongColor,mark=*}}

\newcommand{\MyVec}[1]{\mathbf{#1}}
\newcommand{\Entr}[1]{\ensuremath{\mathbb{H}\left(#1\right)}}

\usepackage{mathtools}
\DeclarePairedDelimiter\abs{\lvert}{\rvert}%
\let\norm\relax
\DeclarePairedDelimiter\norm{\lVert}{\rVert}%
\let\floor\relax
\DeclarePairedDelimiter\floor{\lfloor}{\rfloor}
\makeatletter
\let\oldabs\abs
\def\abs{\@ifstar{\oldabs}{\oldabs*}}
\let\oldnorm\norm
\def\norm{\@ifstar{\oldnorm}{\oldnorm*}}
\let\oldfloor\floor
\def\floor{\@ifstar{\oldfloor}{\oldfloor*}}
\makeatother

\newcommand{\Npermsfun}[1]{\ensuremath{M\!\left( #1 \right)}\xspace}
\newcommand{\floortwo}[1]{\ensuremath{\floor{#1}_2}}
\newcommand{\DMfun}[0]{\ensuremath{f_{\text{DM}}}\xspace}
\newcommand{\DMfunSubscript}[1]{\ensuremath{f_{\text{DM}_{#1}}}\xspace}
\newcommand{\DMfunInverse}[0]{\ensuremath{f^{-1}_{\text{DM}}}\xspace}

\newcommand{\Lout}{\ensuremath{n}\xspace}
\newcommand{\Lin}{\ensuremath{k}\xspace}

\newcommand{\Rateloss}{\ensuremath{R_\text{loss}}\xspace}

\newcommand{\A}{\ensuremath{A}\xspace}

\newcommand{\Composition}{\ensuremath{C}\xspace}

\newcommand{\Seqx}{\ensuremath{\MyVec{x}^n}\xspace}
\newcommand{\AmpOne}{\ensuremath{\alpha}\xspace}
\newcommand{\AmpTwo}{\ensuremath{\beta}\xspace}
\newcommand{\AmpThree}{\ensuremath{\gamma}\xspace}
\newcommand{\AmpFour}{\ensuremath{\delta}\xspace}

\newcommand{\placeholder}{\ensuremath{\underline{\hspace{2mm}}}}

\newcommand{\SetN}[0]{\ensuremath{\mathcal{N}}\xspace}
\newcommand{\SetS}[0]{\ensuremath{\mathcal{S}}\xspace}
\newcommand{\SubsetT}[0]{\ensuremath{\mathcal{T}}\xspace}
\newcommand{\SubsetTLex}[0]{\ensuremath{\overrightarrow{\mathcal{T}}}\xspace}
\newcommand{\SubsetTColex}[0]{\ensuremath{\overleftarrow{\mathcal{T}}}\xspace}
\newcommand{\SubsetTPrime}[0]{\ensuremath{\mathcal{T}^\prime}\xspace}

\newcommand{\rankfun}[0]{\ensuremath{f_{\text{rank}}}\xspace}
\newcommand{\unrankfun}[0]{\ensuremath{f_{\text{unrank}}}\xspace}
\newcommand{\RankFunLex}[0]{\ensuremath{\text{rank}_\text{lex}}\xspace}
\newcommand{\UnrankFunLex}[0]{\ensuremath{\text{unrank}_\text{lex}}\xspace}
\newcommand{\RankFunColex}[0]{\ensuremath{\text{rank}_\text{colex}}\xspace}
\newcommand{\UnrankFunColex}[0]{\ensuremath{\text{unrank}_\text{colex}}\xspace}

\newcommand{\udw}[0]{\ensuremath{\MyVec{b}^\Lin}\xspace} 
\newcommand{\Rank}[0]{\ensuremath{r}\xspace}
\newcommand{\RankLex}[0]{\ensuremath{r_\text{lex}}\xspace}
\newcommand{\RankColex}[0]{\ensuremath{r_\text{colex}}\xspace}
\newcommand{\Weight}[0]{\ensuremath{w}\xspace} 
\newcommand{\BC}[1]{\ensuremath{\text{BC}_{#1}}\xspace}

\usepackage{hyperref} 
\hypersetup{
    colorlinks,
    linkcolor={red!50!black},
    citecolor={blue!50!black},
    urlcolor={blue!80!black}
}
\begin{document}

\title{Parallel-Amplitude Architecture and Subset\\Ranking for Fast Distribution Matching}

\author{Tobias~Fehenberger,~\IEEEmembership{Member,~IEEE,}
        David S. Millar,~\IEEEmembership{Member,~IEEE,} Toshiaki Koike-Akino,~\IEEEmembership{Senior Member,~IEEE,} Keisuke Kojima,~\IEEEmembership{Senior Member,~IEEE,} and Kieran Parsons,~\IEEEmembership{Senior Member,~IEEE}
        
\thanks{T. Fehenberger was with the Technical University of Munich, Munich, Germany, and also with Mitsubishi Electric Research Laboratories, Cambridge,
MA 02139 USA. He is now with the Eindhoven University of Technology. E-mail: tobias.fehenberger@ieee.org.}
\thanks{D. S. Millar, T. Koike-Akino, K. Kojima and K. Parsons are with Mitsubishi Electric Research Laboratories, Cambridge, MA 02139 USA. E-mails: millar@merl.com; koike@merl.com; kojima@merl.com; parsons@merl.com.}}

\markboth{T. Fehenberger \MakeLowercase{\textit{et al.}}: Parallel-Amplitude Architecture and Subset Ranking for Fast Distribution Matching}%
{}

\maketitle

\begin{abstract}
  A distribution matcher (DM) maps a binary input sequence into a block of nonuniformly distributed symbols. To facilitate the implementation of shaped signaling, fast DM solutions with high throughput and low serialism are required. We propose a novel DM architecture with parallel amplitudes (PA-DM) for which $m-1$ component DMs, each with a different binary output alphabet, are operated in parallel in order to generate a shaped sequence with $m$ amplitudes. With negligible rate loss compared to a single nonbinary DM, PA-DM has a parallelization factor that grows linearly with $m$, and the component DMs have reduced output lengths. For such binary-output DMs, a novel constant-composition DM (CCDM) algorithm based on subset ranking (SR) is proposed. We present SR-CCDM algorithms that are serial in the minimum number of occurrences of either binary symbol for mapping and fully parallel in demapping. For distributions that are optimized for the additive white Gaussian noise (AWGN) channel, we numerically show that PA-DM combined with SR-CCDM can reduce the number of sequential processing steps by more than an order of magnitude, while having a rate loss that is comparable to conventional nonbinary CCDM with arithmetic coding.
\end{abstract}


\IEEEpeerreviewmaketitle

\section{Introduction}

  Since its proposal in 2015, probabilistic amplitude shaping (PAS) \cite[Sec.~IV]{Boecherer2015TransComm_ProbShaping} has attracted a lot of attention as method for incorporating probabilistic shaping into bit-interleaved coded modulation (BICM) systems. The reverse concatenation principle of PAS allows to use existing binary forward error correction (FEC) without the need for demapper-decoder iterations at the receiver. PAS enables significant shaping gains and rate adaptivity for a fixed-rate FEC. It has been used in many different communication settings, such as the optical channel \cite{Fehenberger2015OFC_ProbShaping,Renner2017JLT_ShortReachShaping}, in fiber transmission \cite{Buchali2015ECOC_ProbShapingExp}
   and transatlantic field trials \cite{Cho2017OFC_ShapingFieldTrial}, for orthogonal frequency-division multiplexing \cite[Sec.~IV]{Boecherer2017Arxiv_PDM}, and polar coded modulation \cite{Prinz2017SPAWC_PolarCodePAS}.

  The distribution matcher (DM) plays an integral role in the PAS framework as the transmitter-side device for mapping a sequence of uniform data bits into shaped amplitudes. At the receiver, the inverse operation of demapping is carried out. In this paper, we consider block-wise, fixed-length, invertible DMs with binary input. All finite-length DMs suffer from a rate loss that ultimately limits the throughput of a shaped coded modulation system. The rate loss can by decreased by increasing the DM block length, which has the disadvantages of long processing time (and thus latency) and high memory requirements.

  In order to properly characterize and compare different DMs, we differentiate between the DM \emph{system}, describing the general DM architecture and its properties, and the DM \emph{method}, which relates to the actual implementation (e.g., algorithm) of the DM mapping and demapping function. A widely used DM system is based on constant-composition distribution matching (CCDM) \cite[Sec.~III]{Schulte2016TransIT_DistributionMatcher}, and the proposed algorithm to realize CCDM is arithmetic coding \cite[Sec.~IV]{Schulte2016TransIT_DistributionMatcher}.

  For CCDM, each shaped output sequence has the same composition, i.e., the relative frequency of each amplitude within each block is fixed for all possible output sequences. As shown in \cite[Sec.~III-B]{Schulte2016TransIT_DistributionMatcher}, the CCDM rate becomes negligibly small for output lengths beyond approximately 500~symbols.
  Arithmetic coding (AC) was proposed in \cite[Sec.~IV]{Schulte2016TransIT_DistributionMatcher} for an implementation of CCDM. The main drawback of this method is that it is serial in the number of input bits \Lin for mapping and in the sequence length \Lout of the shaped amplitudes for demapping. To the best of our knowledge, there is no constructive CCDM algorithm other than AC, and we refer to it as AC-CCDM.\footnote{The use of lookup tables is not considered because their size and thus hardware requirements are infeasible in practice at an acceptable rate loss.} The serial nature of AC-CCDM in combination with the long blocks required for low rate loss currently make real-time operation of CCDM highly challenging. 
  

  Recently proposed DM systems lift the constant-composition principle, thereby reducing the length that is required for a certain rate loss in comparison to conventional CCDM. In \cite{Fehenberger2018Arxiv_MPDM}, distribution matching via multiset partitioning is proposed. Shell mapping to index the output sequences is proposed in \cite{Schulte2018Arxiv_ShellMapping}. Both techniques are shown to give a block length reduction by approximately a factor of 5 compared to CCDM. The low-complexity DM of \cite{Cho2016ECOC_ShapingNonlinearTolerance} generates two shaped output sequences for each binary input word and chooses the one with less average energy, which implicitly leads to a Gaussian-like distribution. In \cite{GultekinWillems2017PIMRC_EnumerativeShaping}, an enumerative amplitude shaping method is proposed that is based on choosing those sequences in a trellis that have a certain maximum energy. The DM proposed in \cite{Yoshida2017ECOC_Shaping} compares different sequences generated by a mark ratio controller and selects the sequence that has desired properties.

  Parallelization of a nonbinary-alphabet DMs can be achieved with product distribution matching \cite{Boecherer2017Arxiv_PDM} and bit-level distribution matching (BL-DM) \cite{Pikus2017CommLetters_BLDM}. These two independently proposed schemes realize a nonbinary-to-binary transformation by factorizing the nonbinary distribution of $m$ amplitudes into $\log_2 m$ binary component distributions. In the following, we jointly refer to these two proposals as BL-DM since they carry out the same task. For each bit level, one CCDM is then used whose binary outputs are combined to give the desired nonbinary output sequence. In addition to parallelization by a factor $\log_2 m$ compared to a single nonbinary DM, BL-DM can have a smaller rate loss than employing a single nonbinary DM, at the expense of a limited choice of target distributions as they must be product distributions.

  In this paper, a novel distribution matcher with parallel amplitudes (PA-DM) is proposed for which several binary DMs are operated in parallel instead of a single nonbinary DM.\footnote{All considered DMs have binary input, so the distinction between binary and nonbinary alphabets relates to the DM output only.} A binary CCDM is employed for each of the $m-1$ out of $m$ shaped amplitudes, with the alphabet of each binary output subsequence comprising a specific amplitude symbol or the absence thereof. These subsequences are then sequentially combined to generate the desired sequence of shaped nonbinary symbols. In PA-DM, the numbers of parallel DMs grows linearly with $m$, which results in a higher degree of parallelization than for BL-DM, which has a DM per bit level and thus a number of parallel DMs that is logarithmic in $m$.

  We further propose a method for CCDM mapping and demapping via subset ranking (SR) as an alternative to AC-CCDM for binary alphabets. The proposed method is closely related to the enumerative techniques used by Schalkwijk \cite{Schalkwijk1972TransIT_EnumerativeCoding} and Cover \cite{Cover1973TransIT_EnumerativeCoding}. In this paper, we focus on SR implementations that reduce the number of sequential operations as much as possible. In contrast to AC-CCDM, CCDM mapping with the SR algorithm, which we refer to as SR-CCDM, is serial in the smallest number of occurrences of either binary symbol in the output sequence, and demapping via SR-CCDM is fully parallel. For a distribution with $m=8$ shaped amplitudes that is optimized for the additive white Gaussian noise (AWGN) channel, combining PA-DM and SR-CCDM, is numerically shown to give a reduction in serialism of more than an order of magnitude for similar performance as conventional CCDM.

\section{Preliminaries}

\subsection{Notation}
The realizations $a_i,~i \in \{1,\ldots,m\}$ of a random variable \A are drawn from the alphabet $\mathcal{A}$ according to the probability mass function (PMF) $P_{\A}$. Vectors of length $n$ are denoted as $\Seqx=[x_1,\ldots,x_n]$. If the elements of such a vector are binary, e.g., $a$ and $b$, an equivalent notation is $\{a,b\}^n$. Sets are denoted as calligraphic letters, e.g., $\SetN=\{1,\ldots,\Lout\}$.

\subsection{Probabilistic Amplitude Shaping (PAS)}

\begin{figure*}[t]
  \begin{center}
  \includegraphics[width=2\columnwidth]{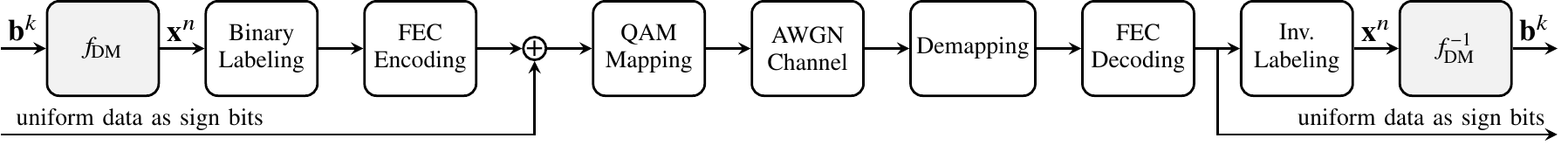}
  \caption{Block diagram of PAS. The plus node combines the shaped amplitude bits (which remain unchanged by the systematic FEC encoder) and the sign bits, which are the parity bits and possible some uniform input bits. This paper covers DM systems and methods (gray boxes).}
  \label{fig:app:PAS_block_diagram}
  \end{center}
\end{figure*}
We briefly outline PAS in the following and refer to \cite[Sec.~IV]{Boecherer2015TransComm_ProbShaping} for details. A block diagram of PAS is shown in Fig.~\ref{fig:app:PAS_block_diagram}, where we assume for simple representation that the DM output and FEC input lengths are compatible. When the DM is shorter than the FEC, several DM sequences must be combined within an FEC block.

The binary data word to be transmitted is split into the DM input sequence \udw and uniform data bits. The DM mapping \DMfun transforms \udw into a sequence \Seqx comprising the shaped amplitudes $\{a_1,,\ldots,a_m\}$. The sequence is given a binary label and input into a systematic FEC encoder. The information bits of the FEC output correspond to the shaped amplitudes, while the parity bits combined with the uniform data bits represent the sign bits of the constellation symbols. After modulation, for instance with two-dimensional quadrature amplitude modulation (QAM), the shaped symbols are transmitted over an channel such as the AWGN channel. The demapper computes log-likelihood ratios that are used for FEC decoding or estimation of the achievable information rate (AIR) for bit-metric decoding (BMD) \cite[Sec.~VI]{Boecherer2015TransComm_ProbShaping}. When decoding is successful, which is assumed herein, the bits that corresponds to shaped amplitude are transformed into the sequence of shaped amplitude bits \Seqx. After DM demapping \DMfunInverse, the initial word \udw is recovered.


\subsection{Constant-Composition Distribution Matching (CCDM)}\label{ssec:prelim_ccdm}
\subsubsection{Principle}
We consider distribution matchers that map a binary input $\udw=\{0,1\}^\Lin$ to a shaped output sequence $\Seqx=[x_1,\ldots,x_n]$ of length \Lout. The DM mapping function establishes an invertible mapping $\DMfun: \udw \rightarrow \Seqx$, and the inverse operation (demapping) is $\DMfunInverse:\Seqx \rightarrow \udw$.

The $m$ different output amplitudes that can occur in \Seqx are taken from the alphabet $\mathcal{A}=\{a_1,\ldots,a_m\}$. The DM output sequence is said to have the composition $\Composition=\{\Lout_1,\ldots,\Lout_{m}\}$ with $\Lout_i$ denoting the number of times the $i$\textsuperscript{th} amplitude $a_i$ occurs, i.e.,
\begin{equation}
  n_i = \abs{\{j: x_j = a_i\}}
\end{equation}
with $i \in \{1,\ldots,m\}$ and $j \in \{1,\ldots,n\}$. This implies that the relative frequency of $a_i$ is $P_{\A}(a_1)=\frac{\Lout_i}{\Lout}$, which is referred to as type \cite[Sec.~II]{Csiszar1998TransIT_Types}. Throughout this paper, the type of all CCDM output sequences is fixed, i.e., all CCDM outputs have the same composition.

\subsection{Input Length and Rate Loss}
The number of input bits \Lin of a DM depends on the number of different output sequences, which is given by the multinomial coefficient
\begin{equation}\label{eq:multinomial_coeff} 
  \Npermsfun{\Composition} = \dbinom{n}{n_1,n_2,\ldots,n_m} = \frac{\left(\sum\limits_{i=1}^{m}n_i\right)!}{\prod\limits_{i=1}^{m}\left(n_i!\right)}.
\end{equation}
It is a natural choice to consider only DMs with an integer number of input bits. The input length \Lin in bits is thus
\begin{equation}\label{eq:DM_input_bits}
  \Lin = \log_2 \floortwo{\Npermsfun{\Composition}},
\end{equation}
where $\floortwo{\cdot}$ denotes rounding down to the closest power of two. The rate loss of a DM is then defined as
\begin{equation}\label{eq:dm_rateloss}
  \Rateloss = \Entr{\A} - \frac{\Lin}{\Lout},
\end{equation}
where $\Entr{\A}$ is the entropy of the quantized amplitudes \A. Such a quantization is necessary in many finite-length cases to achieve an integer-valued composition. The quantization criterion used in this paper is the minimization of Kullback-Leibler divergence between the initial unquantized PMF and the quantized distribution $P_{\A}$ \cite[Sec.~IV]{Boecherer2016TransIT_Quantization}.

\subsubsection{Arithmetic Coding as CCDM Method}
CCDM mapping and demapping can be carried with arithmetic coding (AC) \cite[Sec.~IV]{Schulte2016TransIT_DistributionMatcher}. More details on AC including a discussion of the algorithm implementation can be found in \cite[Ch.~4]{Sayood2012Book_IntroductionDataCompression}. The underlying principle is drawing without replacement where in every AC step, interval boundaries are computed based on those elements of the composition that have not yet been used in the output sequence. Since the size of these intervals depends on previous steps, AC is an inherently sequential algorithm that, in the worst case, is serial in the number of input elements, which is \Lin for mapping and \Lout for mapping.\footnote{Note that there are cases where the AC algorithm can be terminated earlier because the remainder of the output sequence follows with probability 1. We neglect these cases and discuss only worst-case serialism which occurs when all AC steps must be carried out.} Within each serially executed AC operation, the number and complexity of computations to be performed varies as it depends on the specific interval boundaries.

\section{Distribution Matching with Parallel Amplitude Levels}\label{sec:pal_dm}
  In the following, a distribution matching scheme is explained which allows to transform a single DM with nonbinary output into parallel DMs that each have a binary output alphabet corresponding to a shaped amplitude.

  \subsection{Preliminaries: Binomial and Multinomial Coefficients}
  To explain the approach of PA-DM, it is insightful to express the multinomial coefficient $\Npermsfun{\Composition}$ of a composition $\Composition=\{n_1,\ldots,n_m\}$ with length $n=\sum_{i=1}^m n_i$ (see \eqref{eq:multinomial_coeff}) as a product of binomial coefficients (BCs),
  \begin{align}
    \Npermsfun{\Composition} ={}& \dbinom{n}{n_1,n_2,\ldots,n_m} \\
    ={}& \underbrace{\dbinom{n}{n_1}}_{\BC{1}} \cdot \underbrace{\dbinom{n-n_1}{n_2}}_{\BC{2}} \cdot \ldots \cdot \underbrace{\dbinom{n-n_1-\ldots-n_{m-2}}{n_{m-1}}}_{\BC{m-1}} \cdot \underbrace{\dbinom{n_m}{n_{m}}}_{\BC{m}} \label{eq:prod_of_binomials_writtenout} \\
    ={}& \prod\limits_{i=1}^m \underbrace{\dbinom{n-\sum\limits_{j=0}^{i-1}n_j}{n_i}}_{\BC{i}} \label{eq:prod_of_binomials} ,
  \end{align}
  where we define $n_0=0$ in \eqref{eq:prod_of_binomials} for notational convenience. We recall that the first factor of \eqref{eq:prod_of_binomials} represents the number of ways to choose $n_1$ out of $n$ elements (disregarding their order), the second one the ways to choose $n_2$ elements out of the remaining $n-n_1$ elements and so forth. Varying the ordering of the binomial expansion can give different component binomial coefficients (see also Sec.~\ref{ssec:ordering_PA}), but their product is always equal to $\Npermsfun{\Composition}$ and the last factor $\BC{m}$ in \eqref{eq:prod_of_binomials} is equal to 1. In the following, we use the product of binomial coefficients of \eqref{eq:prod_of_binomials} to transform a nonbinary DM into binary component DMs with parallel amplitudes.

  \begin{figure}
  \begin{center}
  \includegraphics[width=\columnwidth]{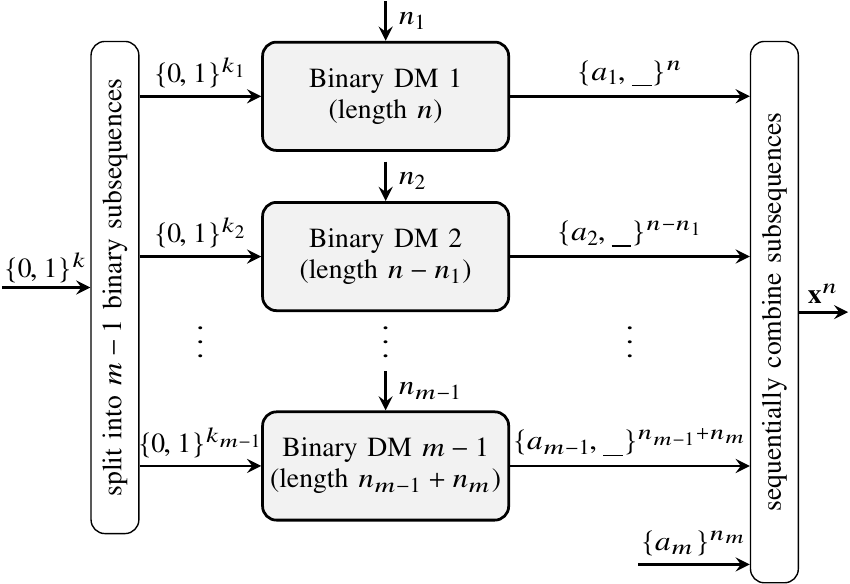}
  \end{center}
  \caption{Mapping structure of distribution matching with parallel amplitudes (PA-DM). $m-1$ parallel binary DMs of varying lengths are employed and their output is sequentially combined to achieve the nonbinary shaped sequence \Seqx.}
  \label{fig:DM_trafo_block_diagram}
  \end{figure}

  \subsection{PA-DM Method}
  In the PA-DM architecture, the first $m-1$ BCs in \eqref{eq:prod_of_binomials} each correspond to a DM instance that maps a binary input to a sequence whose alphabet comprises the considered amplitude and another symbol which we denote as $\placeholder$ and which represents the absence of that amplitude. Figure~\ref{fig:DM_trafo_block_diagram} shows a block diagram of the mapping with PA-DM. First, the binary input of length $\Lin$ is split into $m-1$ substrings that are each input into a binary-input binary-output DM. The mapping operation of the first DM is to place $n_1$ occurrences of the first amplitude $a_1$ in $n$ positions, with the output sequence $\{a_1,\placeholder\}^n$ mapping the input of length $\Lin_1=\floortwo{\BC{1}}$. The second DM maps $\Lin_2=\floortwo{\BC{2}}$ bits by placing the amplitude $a_2$ $n_2$-times in the remaining (unused) $n-n_1$ positions. By repeating this for all amplitudes up to $a_{m-1}$, \Seqx is gradually constructed. Finally, the remaining $n_m$ positions of \Seqx that are not yet occupied are filled with $a_m$. This gives the desired nonbinary output sequence \Seqx.\footnote{We note that this sequential combination can also be done in a tree-like fashion by repeatedly combining two amplitudes at once, which reduces the number of sequential operations.} An example of PA-DM mapping is given below in Example~\ref{ex:product-of-binomials}. Demapping for PA-DM is achieved by performing the above steps in inverse order, i.e., by first decomposing the shaped sequence into binary subsequences, applying inverse distribution matching, and combining the outputs to generate the initially transmitted \udw. The method for the binary component DMs can be either conventional AC-CCDM or the subset-ranking method of Sec.~\ref{sec:dm_sr}.

  An important benefit of PA-DM is that it allows the DM mapping to be split in parallel instances, thereby enabling high-throughput DM implementations. The number of parallel DMs (and thus the parallelization factor compared to a single nonbinary DM) is $m-1$ and hence grows linearly with the one-dimensional modulation order, whereas the number of parallel DMs is logarithmic in $m$ for BL-DM. The parallelization factors compared to a nonbinary DM are summarized in Table~\ref{table:parallelization_PADM_BLDM} for two-sided amplitude shift keying (ASK), i.e., including the PAS sign bit. We observe particularly for high-order modulation that PA-DM employs significantly more DMs than BL-DM. An additional advantage of PA-DM could be that the binary component DMs have decreasing output length and thus computational complexity, while they are of identical length for BL-DM.

\begin{table}
\begin{center}
\caption{Parallelization Factors of PA-DM and BL-DM Compared to Nonbinary DM For Amplitude Shift Keying (ASK) Formats}
\begin{tabular}{c|c||c|c}
 Format & $m$ & PA-DM & BL-DM \\
 \hline
 4ASK (16QAM) & $2$ & 1 & 1\\
 8ASK (64QAM) & $4$ & 3 & 2\\
 16ASK (256QAM) & $8$ & 7 & 3\\
 32ASK (1024QAM) & $16$ & 15 & 4\\
\end{tabular}
\label{table:parallelization_PADM_BLDM}
\end{center}
\end{table}

  \subsection{Rate Loss of PA-DM}
  As each factor of \eqref{eq:prod_of_binomials} corresponds to a DM that maps an integer $\Lin_i=\log_2\floortwo{\BC{i}}$ bits, the aggregate number of input bits of all DMs in the PA-DM architecture is 
\begin{equation}\label{eq:PA_DM_input_bits}
\Lin = \sum_{i=1}^{m} \Lin_i = \sum_{i=1}^{m} \log_2\floortwo{\BC{i}}.
\end{equation}
For a nonbinary DM, we have $\Lin = \floor{\log_2 \Npermsfun{\Composition}}$, see \eqref{eq:DM_input_bits}. Depending on the specific composition, rounding down each individual BC to the largest power of 2 can yield no additional rate loss, or can also result in a small loss up to $m-2$ bits compared to a single nonbinary DM.

  \begin{example}[Mapping Operation for PA-DM]\label{ex:product-of-binomials}
  Consider the composition $\Composition=\{4,3,2,1\}$ for the amplitudes $\{\AmpOne,\AmpTwo,\AmpThree,\AmpFour\}$ and an output sequence \Seqx with $\Lout=10$ that is supposed to have this composition. By \eqref{eq:multinomial_coeff}, we have $\Npermsfun{\Composition} = 12600$ permutations, and thus $\log_2\floortwo{12600}=13$~input bits that can be mapped with a conventional nonbinary DM. By splitting the multinomial coefficient into a product of binomials according to \eqref{eq:prod_of_binomials}, we have $\Npermsfun{\Composition} = \dbinom{10}{4} \cdot \dbinom{6}{3} \cdot \dbinom{3}{2} \cdot \dbinom{1}{1} = 210 \cdot 20 \cdot 3 \cdot 1 = 12600$, which gives $\log_2 \floortwo{210} + \log_2 \floortwo{20} + \log_2 \floortwo{3} + \log_2 \floortwo{1} = 7 + 4 + 1 + 0 = 12$~bits at the PA-DM input. Thus, PA-DM has an additional rate loss of 1 bit compared to a nonbinary DM.
  Now suppose that the 12-bit data word to be mapped is $\udw=[0 1 1 1 0 1 0 0 0 1 0 1]$, which is split into subsequences of lengths $\Lin_1=7$, $\Lin_2=4$, and $\Lin_3=1$.
  Depending on the mapping algorithm (see Sec.~\ref{sec:dm_sr}), the $m-1=3$ DM mapping outputs are as follows, with $\placeholder$ denoting the absence of an amplitude:
  \begin{itemize}
    \item $\DMfunSubscript{1}: [0 1 1 1 0 1 0] \rightarrow [\AmpOne,\placeholder,\placeholder,\AmpOne,\placeholder,\placeholder,\AmpOne,\AmpOne,\placeholder,\placeholder]$
    \item $\DMfunSubscript{2}: [0 0 1 0] \rightarrow [\AmpTwo,\AmpTwo,\placeholder,\placeholder,\AmpTwo,\placeholder]$
    \item $\DMfunSubscript{3}: [1] \rightarrow [\AmpThree,\placeholder,\AmpThree]$
  \end{itemize}
  These three output sequences are then combined sequentially. The 6 free positions of the first DM output are filled with the output of the second DM, giving the temporary sequence $[\AmpOne,\AmpTwo,\AmpTwo,\AmpOne,\placeholder,\placeholder,\AmpOne,\AmpOne,\AmpTwo,\placeholder]$. The remaining 3 positions are used by the third DM and we have $[\AmpOne,\AmpTwo,\AmpTwo,\AmpOne,\AmpThree,\placeholder,\AmpOne,\AmpOne,\AmpTwo,\AmpThree]$. The remaining open position is filled with $\Lout_m=1$ occurrence of $\AmpFour$, giving the final CCDM output sequence $\Seqx = [\AmpOne,\AmpTwo,\AmpTwo,\AmpOne,\AmpThree,\AmpFour,\AmpOne,\AmpOne,\AmpTwo,\AmpThree]$.

  \end{example}

  \subsection{Ordering of Binomial Coefficients}\label{ssec:ordering_PA}

  \begin{figure}
  \begin{center}
  \includegraphics[width=\columnwidth]{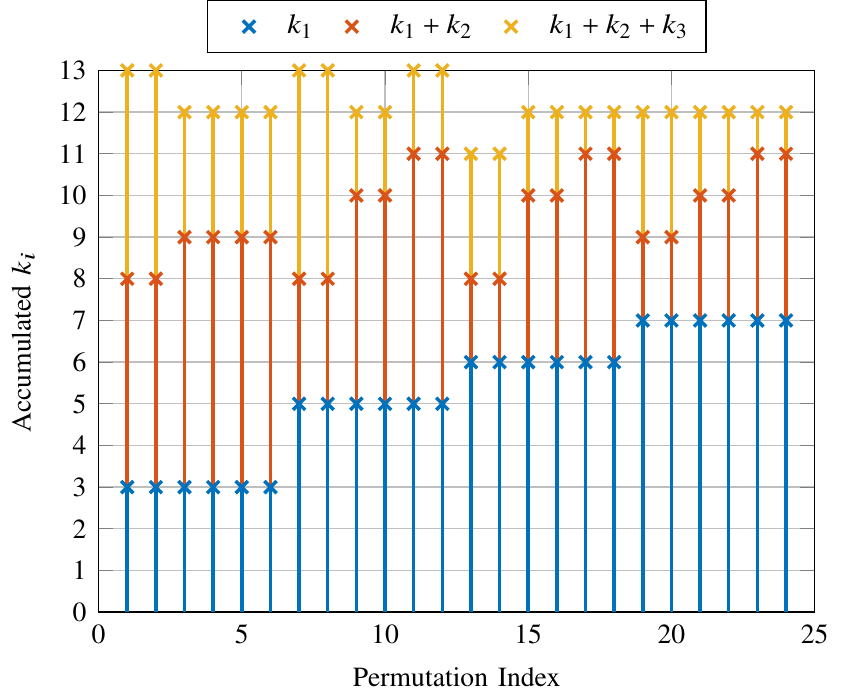}
  \end{center}
  \caption{Accumulated PA-DM input length over permutation index of lexicographically sorted composition $\Composition=\{4,3,2,1\}$. There are six permutations and thus ordering of DMs in the PA-DM system that achieve the maximum input length of 13~bits.}
  \label{fig:BC_ordering_example}
  \end{figure}

  As previously noted in the context of \eqref{eq:prod_of_binomials}, the product of binomial coefficients is always equal to the multinomial coefficient, but the individual factors can vary. Hence, depending on the ordering of the BCs, each component binary DM can take a different number of bits $\Lin_i=\floortwo{\BC{i}}$ at its input. Due to the nonlinearity of the flooring of each BC (see \eqref{eq:PA_DM_input_bits}), the order of the BCs has an impact on $\Lin$ and thus on the DM rate loss. By a simple one-time exhaustive search over all possible orderings (of which there are at most $m!$), the rate loss of PA-DM can be minimized, as illustrated in the following Example~\ref{ex:min_rate_loss}. A detailed rate loss comparison of PA-DM to BL-DM and a nonbinary DM for AWGN-optimized distributions can be found in Sec.~\ref{ssec:num_results:rateloss}.

  \begin{example}[Optimize BC Ordering to Minimize Rate Loss]\label{ex:min_rate_loss}
  Given the composition $\Composition=\{4,3,2,1\}$ for the amplitudes $\{\AmpOne,\AmpTwo,\AmpThree,\AmpFour\}$ as described in Example~\ref{ex:product-of-binomials}, there are $m!=24$ different orderings of the binomial coefficients. In Fig.~\ref{fig:BC_ordering_example}, the accumulated number of input bits $\Lin_i$ per DM is shown over the index of the BC orderings, which are sorted lexicographically. This means that permutation index 1 corresponds to $\Composition=\{1,2,3,4\}$
  and index 24 is $\Composition=\{4,3,2,1\}$. We observe from Fig.~\ref{fig:BC_ordering_example} that there are 6 composition orderings that allow to address $\Lin=13$~bits. One of these orderings is $\Composition=\{1,2,3,4\}$, for which we have $\Npermsfun{\Composition} = \dbinom{10}{1} \cdot \dbinom{9}{2} \cdot \dbinom{7}{3} \cdot \dbinom{4}{4} = 10 \cdot 36 \cdot 35 \cdot 1 = 12600$. In this case $\log_2 \floortwo{10} + \log_2 \floortwo{36} + \log_2 \floortwo{35}=13$~bits can be mapped by PA-DM, resulting in zero additional rate loss in comparison to a single nonbinary DM.
  \end{example}

\section{Distribution Matching via Subset Ranking}\label{sec:dm_sr}
This section outlines binary-output CCDM mapping and demapping methods with low serialism. The key parameters are the output length \Lout in bits, the DM input length defined as $\Lin = \log_2 \floortwo{\dbinom{\Lout}{\Weight}}$, and the weight \Weight denoting the numbers of occurrences of a symbol $a$ in the binary sequence $\Seqx=\{a,b\}^\Lout$, i.e.,
\begin{equation}\label{eq:weight}
  \Weight = \abs{i \in \{1,\ldots,\Lout\}:x_i=a}.
\end{equation}
Since \Seqx is binary, we have $\Lout-\Weight$ occurrences of $b$.

The proposed DM method is based on techniques for the ranking of subsets that are drawn from a set, which is a well-known problem in enumerative combinatorics (e.g. \cite[Sec.~2.4]{Kreher1998Book_CombinatorialAlgorithms}). A similar approach has been applied by Schalkwijk \cite{Schalkwijk1972TransIT_EnumerativeCoding} and Cover \cite{Cover1973TransIT_EnumerativeCoding} for source coding, and recently been used in enumerative sphere shaping
\cite{Willems1993BeNeLux_PragmaticShaping,GultekinWillems2018ISIT_EnumerativeShaping}. We focus on highly parallel algorithms for subset ranking, with an application to CCDM, noting that the proposed approach can be used for any binary enumerative coding technique.

In the following, we review the preliminaries for subset ranking (SR) before linking it to distribution matching. Algorithms are presented, their application is discussed, and compared to a conventional AC-CCDM.

\subsection{Preliminaries and Definitions for Subset Ranking}

Let $\SetN=\{1,\ldots,\Lout\}$ with $\Lout$ being the DM output length as introduced earlier in this manuscript. We further define the set \SetS to consist of the $\dbinom{\Lout}{\Weight}$ $\Weight$-element subsets of the $\Lout$-set \SetN. The $\Weight$-element subset $\SubsetT \subseteq \SetN$ contains the integer elements $\{t_1,\ldots,t_\Weight\}$ and thus constitutes the elements of \SetS.

We are interested in ordering the subsets \SetS, for which a natural choice is lexicographic (lex) ordering. To impose this order on \SetS, we first introduce the list representation \SubsetTLex of \SubsetT as
\begin{equation}\label{eq:list_lex_def}
  \SubsetTLex = \left[t_1,t_2,\ldots,t_\Weight\right],
\end{equation}
where the elements of \SubsetT are sorted in ascending order,
\begin{equation}\label{eq:lex_ordering}
  t_1 \leq t_2 \leq \ldots \leq t_\Weight,
\end{equation}
as indicated by the arrow direction of \SubsetTLex. The lex ordering of the subsets \SetS is obtained by sorting the sequences \SubsetTLex in a dictionary-style fashion, i.e., by applying ascending order to the component integers $\left[t_1,t_2,\ldots,t_\Weight\right]$.

Another common ordering besides lex is colexicographic (colex). In analogy to \eqref{eq:list_lex_def}, we define the colex list representation of \SubsetT as
\begin{equation}\label{eq:list_colex_def}
  \SubsetTColex = \left[t_1,t_2,\ldots,t_\Weight\right].
\end{equation}
with
\begin{equation}\label{eq:colex_ordering}
  t_1 \geq t_2 \geq \ldots \geq t_\Weight.
\end{equation}
The colex ordering in \SetS is achieved by applying lex ordering to all sequences \SubsetTColex.

Given a specific ordering, the \emph{ranking} of a $\Weight$-element subset \SubsetT determines its position (or rank) within all $\dbinom{n}{\Weight}$ subsets \SetS. Thus, the rank of a specific subset for a given ordering is the number of precursors that this subset has. Formally, a ranking is a bijective function from $\SetS$ to the integer rank \Rank, i.e.,
\begin{equation}\label{eq:rank_fun_def}
  \rankfun: \SetS \rightarrow \Rank, \qquad \Rank \in \{0,\ldots,\dbinom{\Lout}{\Weight}-1\}.
\end{equation}
The inverse operation is called \emph{unranking} and defined as
\begin{equation}\label{eq:unrank_fun_def}
  \unrankfun: \Rank \rightarrow \SetS, \qquad \Rank \in \{0,\ldots,\dbinom{\Lout}{\Weight}-1\}.
\end{equation}

In the following, we introduce a specific notation for the ranking and unranking functions depending on the ordering, which is indicated by the subscript lex or colex. The ranking of a particular subset \SubsetT with lex ordering is denoted as 
\begin{equation}\label{eq:rank_fun_lex}
  \RankFunLex\left(\SubsetT\right) = \RankLex,
\end{equation}
and we write for colex ordering
\begin{equation}\label{eq:rank_fun_colex}
  \RankFunColex\left(\SubsetT\right) = \RankColex.
\end{equation}
In analogy, the unranking functions are
\begin{equation}\label{eq:unrank_fun_lex}
  \UnrankFunLex\left(\RankLex\right) = \SubsetT
\end{equation}
for lex and
\begin{equation}\label{eq:unrank_fun_colex}
  \UnrankFunColex\left(\RankColex\right) = \SubsetT,
\end{equation}
for colex. Note that lex and colex ranking are linked with the simple relation \cite[Theorem 2.4]{Kreher1998Book_CombinatorialAlgorithms}
\begin{equation}\label{eq:rank_lex_colex_relation}
  \RankFunLex\left(\SubsetT\right) + \RankFunColex\left(\SubsetTPrime\right) = \dbinom{\Lout}{\Weight}-1,
\end{equation}
where
\begin{equation}\label{eq:rank_lex_colex_relation2}
  \SubsetTPrime = \{\Lout + 1 - t_i: t_i \in \SubsetT\}.
\end{equation}
This relationship can be useful if ranking or unranking algorithms of a certain ordering has computational advantages.

\begin{example}[Ranking for Lex and Colex Ordering]\label{ex:ranking_lex_colex}
  Consider $\Weight=2$ and the $\Lout=5$-set $\SetN=\{1,\ldots,5\}$. There are $\dbinom{5}{2}=10$ subsets \SubsetT in the set \SetS. The subsets, their list representation and the corresponding ranking for lex and colex ordering are listed in Table~\ref{table:lex_colex_example}.
\end{example}
\begin{table}
  \caption{Ranks for lex (left) and colex (right) ordering for $\Lout=5$ and $\Weight=2$ as per Example~\ref{ex:ranking_lex_colex}.}
  \label{table:lex_colex_example}
  \begin{center}
  \begin{tabular}{c|c|c}
    \SubsetT & \SubsetTLex & \RankLex \\
    \hline
    \{1,2\} & \{1,2\} & 0\\
    \{1,3\} & \{1,3\} & 1\\
    \{1,4\} & \{1,4\} & 2\\
    \{1,5\} & \{1,5\} & 3\\
    \{2,3\} & \{2,3\} & 4\\
    \{2,4\} & \{2,4\} & 5\\
    \{2,5\} & \{2,5\} & 6\\
    \{3,4\} & \{3,4\} & 7\\
    \{3,5\} & \{3,5\} & 8\\
    \{4,5\} & \{4,5\} & 9\\
  \end{tabular}
  \hspace{5mm}
\begin{tabular}{c|c|c}
    \SubsetT &\SubsetTColex & \RankColex \\
    \hline
    \{1,2\} & \{2,1\} & 0\\
    \{1,3\} & \{3,1\} & 1\\
    \{2,3\} & \{3,2\} & 2\\
    \{1,4\} & \{4,1\} & 3\\
    \{2,4\} & \{4,2\} & 4\\
    \{3,4\} & \{4,3\} & 5\\
    \{1,5\} & \{5,1\} & 6\\
    \{2,5\} & \{5,2\} & 7\\
    \{3,5\} & \{5,3\} & 8\\
    \{4,5\} & \{5,4\} & 9\\
  \end{tabular}
  \end{center}
\end{table}

\subsection{Binary Sequence as a Constant-Order Subset}\label{ssec:sr:bin_seq}
A binary sequence $\{a,b\}^n$ with alphabet $\{a,b\}$ can be described by a subset of the integers $\{1,\ldots,n\}$ that denotes the positions of either symbol, for example $a$, in that sequence. The complementary set then gives the locations of the other symbol, here $b$. Applying a constant order (such as ascending) to this integer subset gives an equivalent description of the sequence $\{a,b\}^n$. This correspondence is used in the next section to propose a CCDM method via subset ranking.
\begin{example}[Binary Sequence as Integer Subset]
  Suppose we have the binary sequence $\{a a b b a b a a\}$ with $n=8$. The integer subset describing the positions of symbol $a$ in ascending order is $\{1,2,5,7,8\}$. The complementary subset for $b$ is thus $\{3,4,6\}$.
\end{example}

\subsection{Subset Unranking and Ranking as DM Mapping and Demapping}
We now link the above outlined subset ranking to the DM terminology. With the ranking and unranking functions \eqref{eq:rank_fun_lex} to \eqref{eq:unrank_fun_colex}, a bijective mapping between the subset \SubsetT and its rank is established. The rank (in binary representation) of \SubsetT corresponds to the uniform binary sequence \udw that is the input of the binary-alphabet DM mapper. The $\Weight$-element subset \SubsetT that corresponds to this rank describes which positions of the DM mapper output sequence carry one of the two binary output symbols, see Sec.~\ref{ssec:sr:bin_seq}.\footnote{Which symbol is represented by \Weight is a somewhat arbitrary choice. The same SR functionality is achieved when the $\Weight$-element subset \SubsetT represents the positions of the complementary binary symbol.} The DM mapping operation from uniform data word \udw to shaped sequence \Seqx can thus be considered an unranking problem. In analogy, the DM demapper carries out a ranking operation: given a shaped sequence that corresponds to the subset \SubsetT, the rank is to be determined. 

\begin{example}[DM Mapping and Demapping with Lex Subset Ranking]
  Consider a binary DM with $\Lout=10$ and the desired binary distribution $P_{\A}(0)=0.6$, $P_{\A}(1)=0.4$. We have $\Weight=4$ and thus the DM input length $\Lin = \floortwo{\dbinom{10}{4}}=7$~bits. Suppose the binary word to be mapped is $\udw=[1110101]$, which is $\RankLex=117$ in denary representation. With an unranking algorithm of Sec.~\ref{ssec:sr:algo}, the subset \SubsetT in lex ordering that has $\RankLex=117$ is determined as $\SubsetTLex = [2,4,8,9]$.\footnote{Note that colex ordering is also feasible. For $\RankColex=117$ we would get $\SubsetTColex = [9,8,6,3]$.} From this, the DM output sequence of length 10 is determined, as follows in Sec.~\ref{ssec:sr:bin_seq}. The sequence elements that have indices $[2,4,8,9]$ are set to `1', i.e., we have $\Seqx=[0101000110]$. At the demapper, \SubsetTLex is determined from the sequence \Seqx, and a ranking of this subset gives the initial data word $\udw=[1110101]$.
\end{example}

\subsection{Ranking and Unranking Algorithms}\label{ssec:sr:algo}
In the following, we present pseudo-code algorithms for subset ranking and unranking \cite[Sec.~2.4]{Kreher1998Book_CombinatorialAlgorithms} and discuss their serialism.
Ranking for the subset \SubsetT in lex and colex ordering is presented in Algorithms~\ref{alg:lex_rank} and \ref{alg:colex_rank}, respectively. We note that the inner nested for-loop of Algorithm~\ref{alg:lex_rank} (line \ref{alg:lex_rank:nested_loop}) can be easily replaced with parallel vector operations, which makes the lex ranking algorithm serial in \Weight. The colex ranking does not have any loops and is thus of great interest for low-latency high-throughput DM demapping. 

The unranking algorithms for lex and colex ordering are given as Algorithms~\ref{alg:lex_unrank} and \ref{alg:colex_unrank}, respectively. The inner nested loops in Algorithm~\ref{alg:lex_unrank} (line~\ref{alg:lex_unrank:nested_loop}) and Algorithm~\ref{alg:colex_unrank:nested_loop} (line~\ref{alg:colex_unrank:nested_loop}) can again be executed in parallel. Furthermore, if $\Weight>\frac{\Lout}{2}$, the unranking algorithm can be set to determine the positions of the complementary binary symbol, thereby limiting the required number of loop iterations in the unranking Algorithms~\ref{alg:lex_unrank} and~\ref{alg:colex_unrank} to $\min(\Weight,\Lout-\Weight)$.

\begin{algorithm}
  \caption{Lex ranking function $\RankFunLex(\cdot)$ of \eqref{eq:rank_fun_lex}}\label{alg:lex_rank}
  \begin{algorithmic}[1]
  \Require \SubsetTLex, \Weight \Comment{Ordered subset, weight of binary seq.}
  \Function{LexRank}{$\protect\overrightarrow{\mathcal{T}}$, \Weight}
    \State $\RankLex \gets 0$
    \State $t_0 \gets 0$ \Comment{For notational convenience}
    \For{$i$ from 1 to \Weight}
      \If{$t_{i-1} + 1 \leq t_i - 1$}
        \For{$j$ from $t_{i-1} + 1$ to $t_i - 1$} \label{alg:lex_rank:nested_loop}
        \State $\RankLex \gets \RankLex + \dbinom{\Lout-j}{\Weight-i}$
        \EndFor
      \EndIf
    \EndFor
    \State \textbf{return} \RankLex \Comment{See \eqref{eq:rank_fun_lex}}
  \EndFunction    
  \end{algorithmic}
\end{algorithm}

\begin{algorithm}
  \caption{Colex ranking function $\RankFunColex(\cdot)$ of \eqref{eq:rank_fun_colex}}\label{alg:colex_rank}
  \begin{algorithmic}[1]
  \Require \SubsetTColex, \Weight \Comment{Ordered subset, weight of binary seq.}
  \Function{ColexRank}{$\protect\overleftarrow{\mathcal{T}}$, \Weight}
    \State $\MyVec{j} \gets [1,2,\dots,\Weight]$\Comment{Integer list from 1 to \Weight}
    \State $\RankColex \gets \sum\limits_i^{\Weight} \dbinom{t_i -1}{\Weight + 1 - \MyVec{j}_i}$
    \State \textbf{return} \RankColex \Comment{See \eqref{eq:rank_fun_colex}}
  \EndFunction    
  \end{algorithmic}
\end{algorithm}

\begin{algorithm}
  \caption{Lex unranking function $\UnrankFunLex(\cdot)$ of \eqref{eq:unrank_fun_lex}}\label{alg:lex_unrank}
  \begin{algorithmic}[1]
  \Require \Lout,\Weight,\RankLex \Comment{DM output length, weight of binary seq., rank}
  \Function{LexUnrank}{\Lout,\Weight,\RankLex}
    \State $j \leftarrow 1$
    \For{$i$ from 1 to \Weight}
      \While{$\dbinom{\Lout-j}{\Weight-i} \leq \RankLex$} \label{alg:lex_unrank:nested_loop}
        \State $\RankLex \gets \RankLex - \dbinom{\Lout-j}{\Weight-i}$
        \State $j \gets j + 1$
      \EndWhile
      \State $t_i \gets j$
      \State $j \gets j + 1$
    \EndFor
    \State \textbf{return} $\SubsetTLex=\left[t_1,t_2,\ldots,t_\Weight\right]$\Comment{See \eqref{eq:list_lex_def}}
  \EndFunction    
  \end{algorithmic}
\end{algorithm}

\begin{algorithm}
  \caption{Colex unranking function $\UnrankFunColex(\cdot)$ of \eqref{eq:unrank_fun_colex}}\label{alg:colex_unrank}
  \begin{algorithmic}[1]
  \Require \Lout,\Weight,\RankColex \Comment{DM output length, weight of binary seq., rank}
  \Function{ColexUnrank}{\Lout,\Weight,\RankColex}
    \State $j \leftarrow \Lout$
    \For{$i$ from 1 to \Weight}
      \While{$\dbinom{j}{\Weight+1-i} > \RankColex$} \label{alg:colex_unrank:nested_loop}
        \State $j \gets j - 1$
      \EndWhile
      \State $t_i \gets j + 1$
      \State $\RankColex \gets \RankColex - \dbinom{j}{\Weight+1-i} $
    \EndFor
    \State \textbf{return} $\SubsetTColex=\left[t_1,t_2,\ldots,t_\Weight\right]$\Comment{See \eqref{eq:list_colex_def}}
  \EndFunction    
  \end{algorithmic}
\end{algorithm}



\subsection{Comments on Computational Complexity}
We observe from the above algorithms that an integral part of ranking and unranking is to compute binomial coefficients. For the considered application as DM mapping and demapping functions, it is important that the computation is exact since an inaccurate rank calculation, for instance due to rounding, could lead to the DM introducing a transmission error. Thus, integer arithmetic should be employed rather than relying on typical floating-point precision. We further note that the values of the binomial coefficients can be huge for typical DM lengths. For instance, for a short binary CCDM with $\Lout=100$, binomial coefficients must be computed that exceed the maximum value of an unsigned 64-bit integer.

A method of computing binomial coefficients that could be particularly suitable for such large numbers is by prime factorization of $n!$, where $n$ is integer. We first note that only prime numbers $p \leq n$ appear in the factorization of $n$. The number of times that $n!$ is divisible by the prime $p$, which we denote as $d_p\!\left(n!\right)$, is defined as
\begin{equation}\label{eq:legendre_def}
  d_p\!\left(n!\right) = \sum\limits_{i=1}^{\floor{\log_{p} n}} \floor{\frac{n}{p^i}}.
\end{equation}
This expression is known as Legendre's theorem \cite[Sec.~2.6]{Moll2012Book_NumbersFunctions}.
With this relation, the factorial can be expressed as
\begin{equation}\label{eq:factorial_legendre}
  n! = \prod\limits_{p=2}^{n} p^{d_p\!\left(n!\right)},
\end{equation}
where $p$ is prime and the product runs over prime numbers only, i.e., $p \in \{2,3,5,7,\dots\}$.

The above definition of a factorial can be applied to calculating a binomial coefficient $\dbinom{n}{\Weight}$. With the concept of prime factorization, we have
\begin{equation}\label{eq:binom_coeff_legendre}
  \dbinom{n}{\Weight} = \frac{n!}{(n-\Weight)!\cdot \Weight!} = \frac{\prod\limits_{p=2}^{n} p^{d_p\!\left(n!\right)}}{\left(\prod\limits_{p=2}^{n-\Weight} p^{d_p\!\left((n-\Weight)!\right)}\right) \cdot \left(\prod\limits_{p=2}^{\Weight} p^{d_p\!\left(\Weight!\right)}\right)},
\end{equation}
with the products again over primes only. The computations for \eqref{eq:binom_coeff_legendre} can be further simplified by excluding those elements in the numerator and denominator that will eventually cancel out. The definition \eqref{eq:binom_coeff_legendre} can be beneficial because the numbers in intermediate steps of computing the binomial coefficient are relatively small; neither the bases nor exponents of \eqref{eq:binom_coeff_legendre}, i.e., the primes $p$ and $d_p\!\left(n!\right)$ as per \eqref{eq:legendre_def}, exceed $n$. Also, the computation can partly be implemented with bit shifts and additions.
\begin{example}
  We wish to compute $21!$. The relevant primes are $p=\{2,3,5,7,11,13,17,19\}$. With the exponents computed as per \eqref{eq:legendre_def}, we have $21! = 2^{18} \cdot 3^{9} \cdot 5^{4} \cdot 7^{3} \cdot 11^{1} \cdot 13^{1} \cdot 17^{1} \cdot 19^{1} =51090942171709440000$. In particular, the multiplication of the already huge number $3^{9} \cdot 5^{4} \cdot 7^{3} \cdot 11^{1} \cdot 13^{1} \cdot 17^{1} \cdot 19^{1}$ with $2^{18}$ can be performed efficiently with 18 bit shifts.
\end{example}

\section{Numerical Results}
In the following, we compare the finite-length rate loss of the PA-DM of Sec.~\ref{sec:pal_dm} to a nonbinary DM and the BL-DM system of \cite{Boecherer2017Arxiv_PDM,Pikus2017CommLetters_BLDM}. The reduction in serialism from the subset-ranking (SR) CCDM technique of Sec.~\ref{sec:dm_sr} compared to CCDM via arithmetic-coding (AC), denoted as AC-CCDM, is analyzed in Sec.~\ref{ssec:num_results_dos}.

\subsection{Rate Loss Comparison}\label{ssec:num_results:rateloss}
\begin{figure}
\begin{center}
\includegraphics[width=\columnwidth]{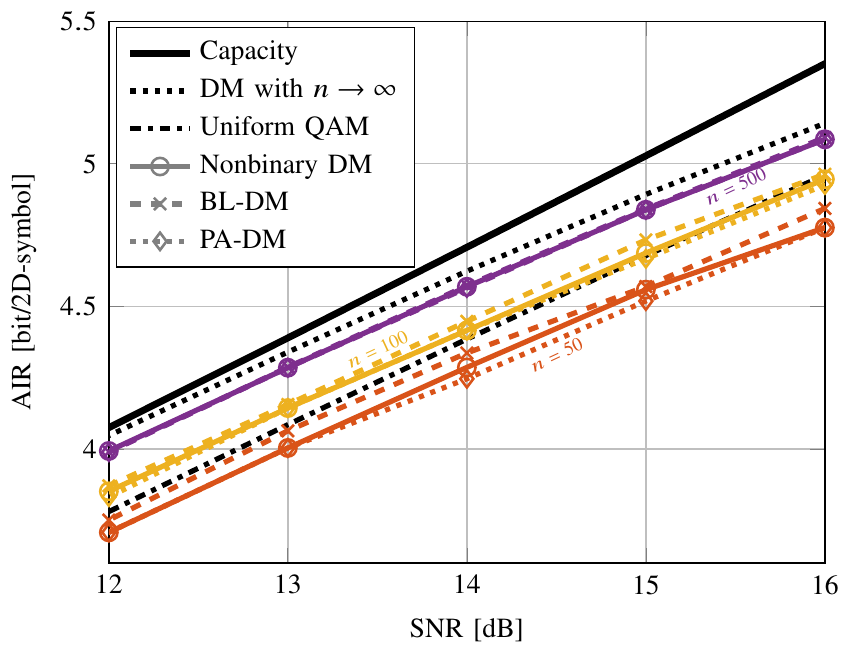}
\end{center}
\caption{Achievable information rate (AIR) for bit metric decoding and finite-length DM systems with $\Lout=\{50,100,500\}$ (colored lines with markers) versus the SNR in dB of the AWGN channel for 64QAM ($m=4$). The channel capacity and the AIRs for an infinite-length DM and for uniform signaling are included as references (black lines).} 
\label{fig:AIR_SNR_PoB_vs_PDM}
\end{figure}

Numerical simulations over the AWGN channel are performed to compare the performance of PA-DM to a nonbinary (NB) DM and BL-DM. The figure of merit is the achievable information rate (AIR) for complex QAM signaling and bit-metric decoding \cite[Sec.~VI]{Boecherer2015TransComm_ProbShaping} minus the finite-length rate loss of the considered DM system, which gives an AIR for the finite-length DM, see \cite[Appendix]{Fehenberger2018Arxiv_MPDM}. The AIRs for 64QAM as a function of the signal-to-noise ratio (SNR) of the AWGN channel are shown in Fig.~\ref{fig:AIR_SNR_PoB_vs_PDM} for PA-DM (dotted), BL-DM (dashed), and conventional nonbinary DM (solid). The channel capacity $\log_2(1+\text{SNR})$ and the asymptotic AIR for infinite-length DM (i.e., with zero rate loss) and uniform signaling are included for reference. The targeted PMF is the optimal Maxwell-Boltzmann PMF \cite{Kschischang1993TransIT_Shaping} at each SNR, quantized for each block length \Lout as to minimize Kullback-Leibler divergence \cite[Sec.~IV]{Boecherer2016TransIT_Quantization}. We observe from Fig.~\ref{fig:AIR_SNR_PoB_vs_PDM} that for short lengths such as $\Lout=50$, BL-DM has improved performance over NB-DM and PA-DM. The reason for this is that the sum of rate losses of the individual BL-DM instances is smaller than the total rate loss of the other schemes. Note, however, that this length regime is of limited interested since the throughput is smaller than with uniform 64QAM. The performance improvement of BL-DM over the other DM systems decreases with increasing \Lout. For $\Lout=500$~symbols, all three investigated systems have nearly identical performance. We further note that the rate loss of PA-DM is smaller than 0.05~bits/2D-sym compared a single nonbinary DM for all considered output lengths and SNRs.

In the following, we perform a detailed analysis of the DM systems for 64QAM, $\Lout=100$ and 13~dB SNR. The results are listed in Table~\ref{table:comparison_pbc_pdm}. First and foremost, we note that the rate losses are very similar: 0.09~bits per 1D amplitude symbol for BL-DM and 0.1~bit for the NB-DM and PA-DM. The parameters of the individual binary DMs are also given in Table~\ref{table:comparison_pbc_pdm}. In comparison to BL-DM, PA-DM uses three binary DMs instead of two, thus allowing a higher degree of parallelization, see also Table~\ref{table:parallelization_PADM_BLDM}. Furthermore, the output lengths \Lout, number of input bits \Lin and smallest number of occurrences  \Weight of either binary symbol is smaller for the component DMs of PA-DM compared to BL-DM, which potentially allows a DM implementation with a smaller number of sequential computations. As the reduction in serialism depends on the employed algorithm, we compare in the following the degree of serialism between AC-CCDM outlined in Sec.~\ref{ssec:prelim_ccdm} and SR-CCDM introduced in Sec.~\ref{ssec:sr:algo}.


\begin{table}
\begin{center}
\caption{Comparison of PA-DM, BL-DM, and nonbinary (NB) DM, all for $\Lout=100$ and 64QAM ($m=4$) at 13~dB SNR}
\begin{tabular}{c|c|c|c}
 ~ & PA-DM & BL-DM & NB-DM\\
 \hline
 Number of DMs & $m-1=3$ & $\log_2 m=2$ & 1 \\
 \hline
  Compositions & $(16,32,6,46)$ & \makecell{$(78,22)$ \\ and \\ $(61,39)$} & $(46,32,16,6)$\\
 \hline
 $(\Lout,\Lin,\Weight)$ per DM & \makecell{$(100,60,16)$ \\ $(84,77,32)$ \\ $(52,24,6)$} & \makecell{$(100,72,22)$ \\ $(100,92,39)$} & $(100,161,-)$\\
 \hline
 Total \Rateloss & 0.1 & 0.09 & 0.1 \\
\end{tabular}
\label{table:comparison_pbc_pdm}
\end{center}
\end{table}

\subsection{Degree of Serialism (DoS) Comparison}\label{ssec:num_results_dos}
In order to assess the computational complexity of DM algorithms, we introduce the notion of degree of serialism (DoS), which describes the number of loop iterations that is executed in either scheme for mapping and demapping. Although this metric does not incorporate the complexity or the required number of clock cycles for the operations within each iteration, it can serve as an insightful metric for evaluating the latency and the potential of parallelization for the investigated CCDM algorithms.

\begin{table}
\begin{center}
\caption{Overview of Degree of Serialism (DoS) for SR-CCDM and AC-CCDM}
\begin{tabular}{c|c|c}
 ~ & SR-CCDM & AC-CCDM \\
 \hline
 Mapping & $\min(\Weight,\Lout-\Weight)$ & \Lin \\
 \hline
 Demapping & 1 (no serialism) & \Lout \\
\end{tabular}
\label{table:serialism_AC_SR}
\end{center}
\end{table}

For SR-CCDM, the unranking algorithms have a serialism of $\min(\Weight,\Lout-\Weight)$, and ranking with colex does not require any iterations (see Algorithm~\ref{alg:colex_rank}), which we define as a serialism of 1.\footnote{The serial combination of the component subsequences is neglected here.} In contrast, the AC-CCDM mapping and demapping algorithms are in the worst case serial in the length of their respective inputs, which is \Lin for mapping and \Lout for demapping. This DoS is summarized in Table~\ref{table:serialism_AC_SR}. The combined reduction in DoS from SR-CCDM (with colex sorting) over AC-CCDM is thus
\begin{equation}\label{eq:dos_reduction}
\frac{\Lin + \Lout}{\min(\Weight,\Lout-\Weight)+1}.
\end{equation}
For comparing parallel DM architectures schemes such as PA-DM and BL-DM, the DoS reduction is computed for the respective worst-case component DM.

\begin{figure}
\begin{center}
\includegraphics[width=\columnwidth]{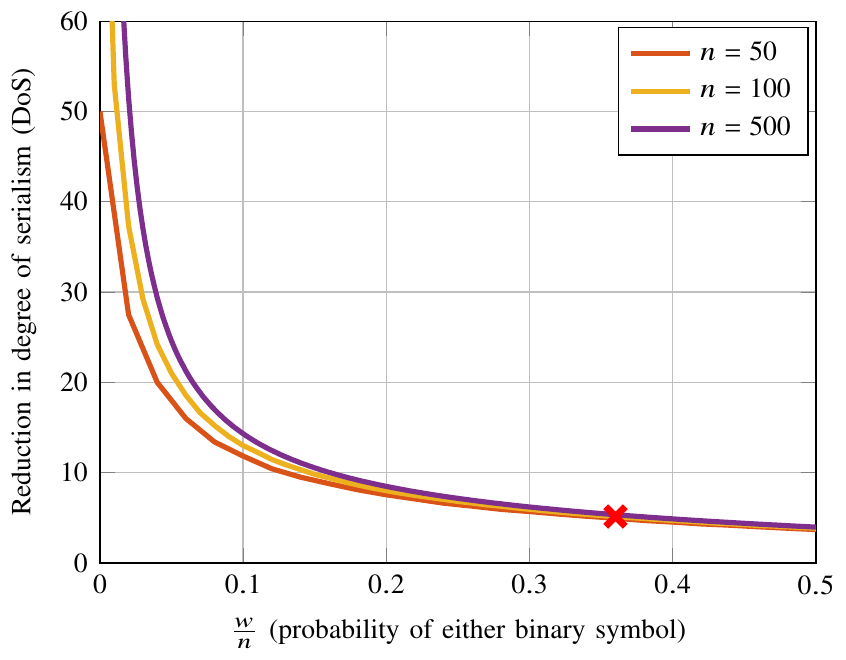}
\end{center}
\caption{Reduction in degree of serialism (DoS) of SR-CCDM compared to AC-CCDM (see \eqref{eq:dos_reduction} for the formal definition) versus the ratio between weight $\Weight$ of one of the binary symbols and DM length, which corresponds to the PMF for that binary symbol. The DoS reduction is shown only for $\Weight \leq \frac{\Lout}{2}$ as the results for larger values of $\Weight$ are a mirrored copy of those presented in the above figure. The marker at $\frac{\Weight}{\Lout}=0.36$ corresponds to Example~\ref{ex:DoS_SR_vs_AC}.} 
\label{fig:Reduction_DoS_BinaryPMF}
\end{figure}

In Fig.~\ref{fig:Reduction_DoS_BinaryPMF}, the DoS reduction is numerically evaluated over $\frac{\Weight}{\Lout}$, which corresponds to the probability of occurrence of either binary symbol, for a CCDM with $\Lout=\{50,100,500\}$ shaped bits out and $\Lin=\log_2\floortwo{\dbinom{\Lout}{\Weight}}$ input bits. We observe that the stronger the binary PMF is shaped, the larger the DoS reduction, which can be more than an order of magnitude for a strongly shaped distribution. 
The following example illustrates the steps of this analysis for $\Weight=64$.

\begin{example}[Serialism of Subset Ranking vs. Arithmetic Coding]\label{ex:DoS_SR_vs_AC}
  Consider a CCDM with $\Lout=100$ and $\Weight=64$, which has $\Lin=90$ input bits. The combined worst-case serialism of AC mapping and demapping is $\Lin+\Lout=190$. Mapping with SR has a serialism of $\min(\Weight,\Lout-\Weight)=36$, and demapping always has serialism~$1$ for colex ordering. Thus, the total reduction in serialism from the subset-ranking method is $190/37\approx5.14$. This reduction is shown in Fig.~\ref{fig:Reduction_DoS_BinaryPMF} as marker. 
\end{example}

Considering the example of Table~\ref{table:comparison_pbc_pdm}, we note that SR-CCDM is also beneficial for the BL-DM system, reducing the DoS of the worst component DM by a factor of 4.8, from $100+92=192$ to $39+1=40$. When using PA-DM instead of BL-DM, the serialism is further reduced to $32+1=33$, corresponding to an improvement of a  factor of 5.8 from SR-CCDM. Compared to a nonbinary DM, the total reduction in serialism from jointly applying PA-DM and SR-CCDM, which is referred to as PASR-CCDM, is $261/33\approx7.9$, at no performance loss.

\begin{figure}
\begin{center}
\includegraphics[width=\columnwidth]{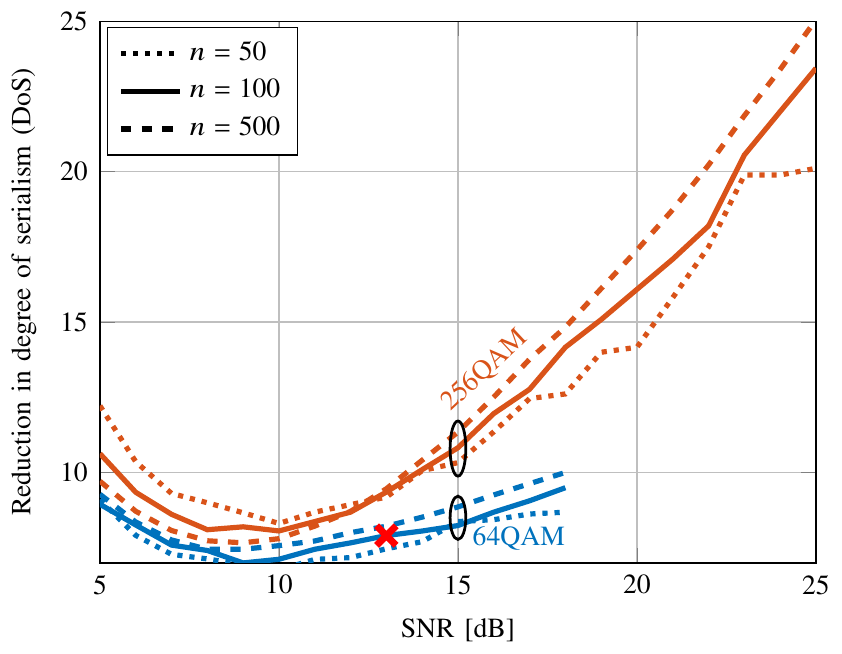}
\end{center}
\caption{Reduction in degree of serialism (DoS) of PASR-CCDM compared to AC-CCDM over the SNR in dB of an AWGN channel for 64QAM and 256QAM. The marker corresponds to Table~\ref{table:comparison_pbc_pdm}.} 
\label{fig:Reduction_DoS_QAM_AWGN}
\end{figure}

In Fig.~\ref{fig:Reduction_DoS_QAM_AWGN}, the reduction in DoS, again for the worst-case DM, is shown for 64QAM and 256QAM over the SNR of the AWGN channel. In analogy to the results of Fig.~\ref{fig:AIR_SNR_PoB_vs_PDM}, the DM compositions are obtained from Maxwell-Boltzmann distributions. We observe that the DoS reduction can be up to a factor 10 for 64QAM, and amount to more than 20 for 256QAM. The additional PASR-CCDM rate loss compared to NB-DM was in all cases either zero or $1/\Lout$, i.e., one extra bit. The reason for the parabola-like shape of the curves as follows. The DoS of NB-DM grows with SNR because for higher SNR, the distribution is more uniform-like, which in general gives a larger $\Lin$ and thus a higher DoS. For PASR-CCDM, however, the DoS depends on each composition (and its ordering), and for the considered compositions, the DoS is numerically found to grow fast at low SNR, causing the dip in the DoS reduction curve, while for high SNR, the DoS of PASR-CCDM grows slower than that of NB-DM. 

\section{Conclusion}
A DM system with parallel amplitudes (PA-DM) has been proposed that employs binary-alphabet DMs for $m-1$ out of $m$ amplitudes (the last amplitude requires no DM). The system has no or negligibly small additional rate loss compared to a single nonbinary DM. The output lengths of the component DMs are decreasing and the number of parallel DMs grows linearly with the modulation order. These features could greatly help to increase the throughput of practical DMs.

We have further introduced a binary-alphabet CCDM mapping and demapping method via subset ranking (SR). A key difference of SR to CCDM via arithmetic coding is that the total number of serial operations required for SR mapping and demapping is the smallest number of occurrences of either binary output symbol (i.e., the minimum weight) plus one. For SR-CCDM, the computational complexity is mostly the calculation of binomial coefficients. Combining PA-DM and SR-CCDM is numerically shown for AWGN-optimized distributions to give a serialism reduction by more than an order of magnitude compared to a nonbinary DM, which could facilitate a practical implementation of short-length CCDMs.

\section*{Acknowledgments}
T.~Fehenberger would like to thank Patrick Schulte (Technical University of Munich) for insightful discussions on the serialism of arithmetic-coding CCDM.

\bibliographystyle{IEEEtran}
\bibliography{Paper_CCDM_SubsetRanking}

\begin{thebibliography}{10}
\providecommand{\url}[1]{#1}
\csname url@samestyle\endcsname
\providecommand{\newblock}{\relax}
\providecommand{\bibinfo}[2]{#2}
\providecommand{\BIBentrySTDinterwordspacing}{\spaceskip=0pt\relax}
\providecommand{\BIBentryALTinterwordstretchfactor}{4}
\providecommand{\BIBentryALTinterwordspacing}{\spaceskip=\fontdimen2\font plus
\BIBentryALTinterwordstretchfactor\fontdimen3\font minus
  \fontdimen4\font\relax}
\providecommand{\BIBforeignlanguage}[2]{{%
\expandafter\ifx\csname l@#1\endcsname\relax
\typeout{** WARNING: IEEEtran.bst: No hyphenation pattern has been}%
\typeout{** loaded for the language `#1'. Using the pattern for}%
\typeout{** the default language instead.}%
\else
\language=\csname l@#1\endcsname
\fi
#2}}
\providecommand{\BIBdecl}{\relax}
\BIBdecl

\bibitem{Boecherer2015TransComm_ProbShaping}
G.~B{\"{o}}cherer, P.~Schulte, and F.~Steiner, ``{Bandwidth efficient and
  rate-matched low-density parity-check coded modulation},'' \emph{{IEEE}
  Transactions on Communications}, vol.~63, no.~12, pp. 4651--4665, Dec. 2015.

\bibitem{Fehenberger2015OFC_ProbShaping}
T.~Fehenberger, G.~B{\"o}cherer, A.~Alvarado, and N.~Hanik, ``{LDPC coded
  modulation with probabilistic shaping for optical fiber systems},'' in
  \emph{Proc. Optical Fiber Communication Conference (OFC)}.\hskip 1em plus
  0.5em minus 0.4em\relax Los Angeles, CA, USA: {Paper Th.2.A.23}, Mar. 2015.

\bibitem{Renner2017JLT_ShortReachShaping}
J.~Renner, T.~Fehenberger, M.~P. Yankov, F.~{Da Ros}, S.~Forchhammer,
  G.~Böcherer, and N.~Hanik, ``{Experimental comparison of probabilistic
  shaping methods for unrepeated fiber transmission},'' \emph{Journal of
  Lightwave Technology}, vol.~35, no.~22, pp. 4871--4879, Nov. 2017.

\bibitem{Buchali2015ECOC_ProbShapingExp}
F.~Buchali, G.~B{\"{o}}cherer, W.~Idler, L.~Schmalen, P.~Schulte, and
  F.~Steiner, ``{Experimental demonstration of capacity increase and
  rate-adaptation by probabilistically shaped 64-QAM},'' in \emph{Proc.
  European Conference and Exhibition on Optical Communication (ECOC)}.\hskip
  1em plus 0.5em minus 0.4em\relax Valencia, Spain: {Paper PDP.3.4}, Sep. 2015.

\bibitem{Cho2017OFC_ShapingFieldTrial}
J.~Cho, X.~Chen, S.~Chandrasekhar, G.~Raybon, R.~Dar, L.~Schmalen, E.~Burrows,
  A.~Adamiecki, S.~Corteselli, Y.~Pan \emph{et~al.}, ``{Trans-atlantic field
  trial using probabilistically shaped 64-QAM at high spectral efficiencies and
  single-carrier real-time 250-Gb/s 16-QAM},'' in \emph{Proc. Optical Fiber
  Communication Conference (OFC)}.\hskip 1em plus 0.5em minus 0.4em\relax Los
  Angeles, CA, USA: {Paper Th5B.3}, Mar. 2017.

\bibitem{Boecherer2017Arxiv_PDM}
G.~B{\"o}cherer, P.~Schulte, and F.~Steiner, ``High throughput probabilistic
  shaping with product distribution matching,'' \emph{arXiv preprint
  arXiv:1702.07510}, Feb. 2017.

\bibitem{Prinz2017SPAWC_PolarCodePAS}
T.~Prinz, P.~Yuan, G.~B{\"o}cherer, F.~Steiner, O.~{\.I}{\c{s}}can,
  R.~B{\"o}hnke, and W.~Xu, ``Polar coded probabilistic amplitude shaping for
  short packets,'' in \emph{Signal Processing Advances in Wireless
  Communications (SPAWC)}, Sapporo, Japan, Jul. 2017.

\bibitem{Schulte2016TransIT_DistributionMatcher}
P.~Schulte and G.~B{\"{o}}cherer, ``{Constant composition distribution
  matching},'' \emph{{IEEE} Transactions on Information Theory}, vol.~62,
  no.~1, pp. 430--434, Jan. 2016.

\bibitem{Fehenberger2018Arxiv_MPDM}
T.~Fehenberger, D.~S. Millar, T.~Koike-Akino, K.~Kojima, and K.~Parsons,
  ``{Multiset-Partition Distribution Matching},'' \emph{{IEEE} Transactions on
  Communications}, 2018.

\bibitem{Schulte2018Arxiv_ShellMapping}
P.~Schulte and F.~Steiner, ``{Shell mapping for distribution matching},''
  \emph{arXiv preprint arXiv:1803.03614}, Mar. 2018.

\bibitem{Cho2016ECOC_ShapingNonlinearTolerance}
J.~Cho, S.~Chandrasekhar, R.~Dar, and P.~J. Winzer, ``{Low-complexity shaping
  for enhanced nonlinearity tolerance},'' in \emph{Proc. European Conference on
  Optical Communications (ECOC)}.\hskip 1em plus 0.5em minus 0.4em\relax
  Düsseldorf, Germany: {Paper W.1.C.2}, Sep. 2016.

\bibitem{GultekinWillems2017PIMRC_EnumerativeShaping}
Y.~C. G{\"u}ltekin, W.~van Houtum, S.~Serbetli, and F.~M. Willems,
  ``{Constellation shaping for IEEE 802.11},'' in \emph{Symposium on Personal,
  Indoor, and Mobile Radio Communications (PIMRC)}, Montreal, QB, Canada, Oct.
  2017.

\bibitem{Yoshida2017ECOC_Shaping}
T.~Yoshida, M.~Karlsson, and E.~Agrell, ``Short-block-length shaping by simple
  mark ratio controllers for granular and wide-range spectral efficiencies,''
  in \emph{Proc. European Conference on Optical Communications (ECOC)}.\hskip
  1em plus 0.5em minus 0.4em\relax Gothenburg, Sweden: {Paper Tu.2.D.2}, Sep.
  2017.

\bibitem{Pikus2017CommLetters_BLDM}
M.~Pikus and W.~Xu, ``{Bit-level probabilistically shaped coded modulation},''
  \emph{{IEEE} Communications Letters}, vol.~21, no.~9, pp. 1929--1932, Sep.
  2017.

\bibitem{Schalkwijk1972TransIT_EnumerativeCoding}
J.~P.~M. Schalkwijk, ``An algorithm for source coding,'' \emph{{IEEE}
  Transactions on Information Theory}, vol.~18, no.~3, pp. 395--399, May 1972.

\bibitem{Cover1973TransIT_EnumerativeCoding}
T.~Cover, ``Enumerative source encoding,'' \emph{{IEEE} Transactions on
  Information Theory}, vol.~19, no.~1, pp. 73--77, Jan. 1973.

\bibitem{Csiszar1998TransIT_Types}
I.~Csisz{\'a}r, ``The method of types,'' \emph{{IEEE} Transactions on
  Information Theory}, vol.~44, no.~6, pp. 2505--2523, Oct. 1998.

\bibitem{Boecherer2016TransIT_Quantization}
G.~B{\"o}cherer and B.~C. Geiger, ``Optimal quantization for distribution
  synthesis,'' \emph{{IEEE} Transactions on Information Theory}, vol.~62,
  no.~11, pp. 6162--6172, Sep. 2016.

\bibitem{Sayood2012Book_IntroductionDataCompression}
K.~Sayood, \emph{{Introduction to data compression}}, 4th~ed.\hskip 1em plus
  0.5em minus 0.4em\relax Elsevier Science, 2012.

\bibitem{Kreher1998Book_CombinatorialAlgorithms}
D.~L. Kreher and D.~R. Stinson, \emph{Combinatorial algorithms: generation,
  enumeration, and search}.\hskip 1em plus 0.5em minus 0.4em\relax CRC press,
  1998, vol.~7.

\bibitem{Willems1993BeNeLux_PragmaticShaping}
F.~M.~J. Willems and J.~J. Wuijts, ``{A pragmatic approach to shaped coded
  modulation},'' in \emph{Symposium on Communications and Vehicular Technology
  in the Benelux}, 1993.

\bibitem{GultekinWillems2018ISIT_EnumerativeShaping}
Y.~C. G{\"u}ltekin, F.~M. Willems, W.~van Houtum, and S.~Serbetli,
  ``{Approximate enumerative sphere shaping},'' in \emph{Proc. IEEE
  International Symposium on Information Theory (ISIT)}, Vail, CO, USA, Jun.
  2018.

\bibitem{Moll2012Book_NumbersFunctions}
V.~Moll, \emph{Numbers and Functions: From a Classical-experimental
  Mathematician's Point of View}.\hskip 1em plus 0.5em minus 0.4em\relax
  American Mathematical Society, 2012.

\bibitem{Kschischang1993TransIT_Shaping}
F.~R. Kschischang and S.~Pasupathy, ``{Optimal nonuniform signaling for
  Gaussian channels},'' \emph{{IEEE} Transactions on Information Theory},
  vol.~39, no.~3, pp. 913--929, May 1993.

\end{thebibliography}

\end{document}